\documentclass{nature}
\usepackage{graphicx} 
\usepackage{isomath}
\usepackage[utf8]{inputenc} 
\usepackage{textcomp}

\newcommand{\squeezeup}{\vspace{-25mm}}
\title{The domination of Saturn's low latitude ionosphere by ring `rain'}
\author{J. O'Donoghue$^1$, T.S. Stallard$^1$, H. Melin$^1$, G.H. Jones$^{2,3}$, S.W.H. Cowley$^1$, S. Miller$^{3,4}$, K.H. Baines$^5$, J.S.D. Blake$^1$} 

\begin{document}
\maketitle

\begin{affiliations}
\footnotesize{ \item Department of Physics and Astronomy, University of Leicester, Leicester LE1 7RH, UK.

  \item Mullard Space Science Laboratory, University College London, Holmbury St. Mary, Dorking, Surrey RH5 6NT, UK.
  
  \item The Centre for Planetary Sciences at UCL/Birkbeck, Gower Street, London WC1E 6BT, UK.
   
  \item Atmospheric Physics Laboratory, Department of Physics and Astronomy, UCL, Gower Street, London WC1E 6BT, UK.
    
  \item Jet Propulsion Laboratory, California Institute of Technology, M/S 183-601, 4800 Oak Grove Drive, Pasadena, CA 91109, USA. }

\end{affiliations}

\maketitle
\textbf{Saturn's ionosphere is produced when the otherwise neutral atmosphere is exposed to a flow of energetic charged particles or solar radiation\cite{tssras}. At low latitudes the latter should result in a weak planet-wide glow in infrared (IR), corresponding to the planet's uniform illumination by the Sun\cite{2010mill}. The observed low-latitude ionospheric electron density is lower and the temperature higher than predicted by models\cite{1984book,2007crisis,2010moore}. A planet-ring magnetic connection has been previously suggested in which an influx of water from the rings could explain the lower than expected electron densities in Saturn's atmosphere\cite{1984conn,1986conn,wilsonring}. Here we report the detection of a pattern of features, extending across a broad latitude band from $\sim$25$^\circ$ to 60$^\circ$, that is superposed on the lower latitude background glow, with peaks in emission that map along the planet's magnetic field lines to gaps in Saturn's rings. This pattern implies the transfer of charged water products from the ring-plane to the ionosphere, revealing the influx on a global scale, flooding between 30 to 43\% of the planet's upper-atmospheric surface. This ring `rain' plays a fundamental role in modulating ionospheric emissions and suppressing electron densities.}

On 17 April 2011 over 2 hours of Saturn near-IR spectral data were obtained by the 10-metre W.M. Keck II telescope using the NIRSPEC spectrometer\cite{nirs}. The slit on the spectrometer was positioned along Saturn's noon meridian as shown in Fig. 1 and the intensity of two bright H$_3^+$ ro-vibrational emission lines is visible almost completely from pole-to-pole, such that low-latitude emissions can be studied. Far from being featureless as we might expect by analogy to Jupiter\cite{tssras} (see also supplementary information), the mid- to low-latitude H$_3^+$ emissions show a number of peaks and troughs before increasing strongly towards the two polar auroral regions; a number of these peaks are observed in both spectral lines. The Q(1, 0$^-$) line shows more substantial peaks and troughs at mid- to low-latitudes than that of the R(2, 2$^-$) line, owing to less contamination by reflected sunlight in neighbouring spectral pixels where methane is not absorbing this light effectively. The apparently symmetric peaks and troughs do not occur at the same latitudes either side of the equator, however, occurring at higher latitudes in the north than in the south. The lack of latitudinal symmetry along with the absence of a similar variability at Jupiter\cite{tssras}, suggests the phenomenon is unrelated to weather patterns or other processes produced in the neutral atmosphere. Instead, the peaks in emission are found to be mapped via planetary magnetic field lines to gaps in Saturn's rings such as the Cassini division (in which we will henceforth include the Herschel, Laplace, Encke and Keeler gaps) and the Colombo gap. In addition, we define the `instability region' as the region between the inner edge of the B-ring and the `instability radius', which are two regions in which the outward centrifugal forces on particles balance with the inward gravitational forces within the rings, such that particles are unstable and can easily stream along magnetic field lines and enter the planet's atmosphere\cite{1982hill,1983hill}. The model used for this mapping employs the most recent internal field coefficients determined from Cassini data\cite{burt2010}, together with small field perturbations produced by magnetospheric currents (see also supplementary information). \\

\begin{figure}
\squeezeup
\squeezeup

\includegraphics[width=35pc]{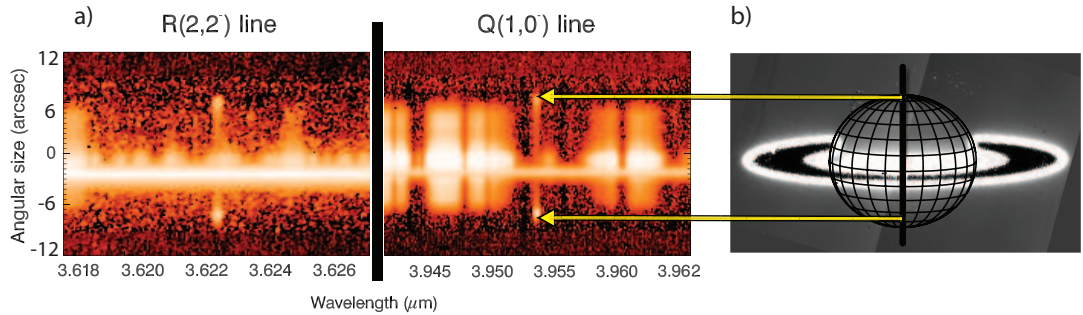}
\caption{{\textbf{The process of data acquisition.} The spectral images shown in a) are separated by a thick black line to indicate the different wavelength ranges. The horizontal and vertical axes in these ranges show wavelength and spatial position along the slit, respectively, whilst intensity ranges from low to high from black to white. Two spectral lines, Q(1, 0$^-$) at 3.953 $\hbox{\textmu}$m and R(2, 2$^-$) at 3.622 $\hbox{\textmu}$m, are centered in each wavelength range, and are from the Q- and R-branches of the H$_3^+$ emission spectrum. These spectra are obtained through the slit of the spectrometer seen in b), which was orientated in the north-south position on Saturn, aligned along the rotational axis. Saturn's spin axis was tilted by 8.2$^{\circ}$ towards Earth during conditions of Saturn early northern spring. The planet rotates beneath the slit, allowing the acquisition of spectral images at a fixed local time, but varying in Saturn longitude. In the $\sim$2 hours of recorded data, 21\% of longitude of the planet was observed - between 101-177$^\circ$ longitude. The bright IR emission measured at the -3 arcsecond position in a) across the entire wavelength range is the uniform reflection of sunlight by the rings, whilst the remaining bright white areas are due to methane reflection. This background methane reflection attenuates the R(2, 2$^-$) line emission more than the Q(1, 0$^-$) line, leading to a lower signal-to-noise ratio.}}
\end{figure}

The field lines within the equatorial region, where the H$_3^+$ emission features are found, map to Saturn's main ring system between 1.2 and 2.3 R$_\mathrm{S}$ from the centre of the planet, as shown by the mapped equatorial distances in Fig. 2. A majority of the emission peaks correspond to prominent gaps in the rings, whilst the troughs map to the dense sections of the rings. This relationship is seen in greater spatial detail in Fig. 3 where ring transparency (as measured by Voyager 2) is compared with the total average H$_3^+$ intensity from the co-addition of both hemispheres and spectral lines. \\

\begin{figure}
\includegraphics[width=35pc]{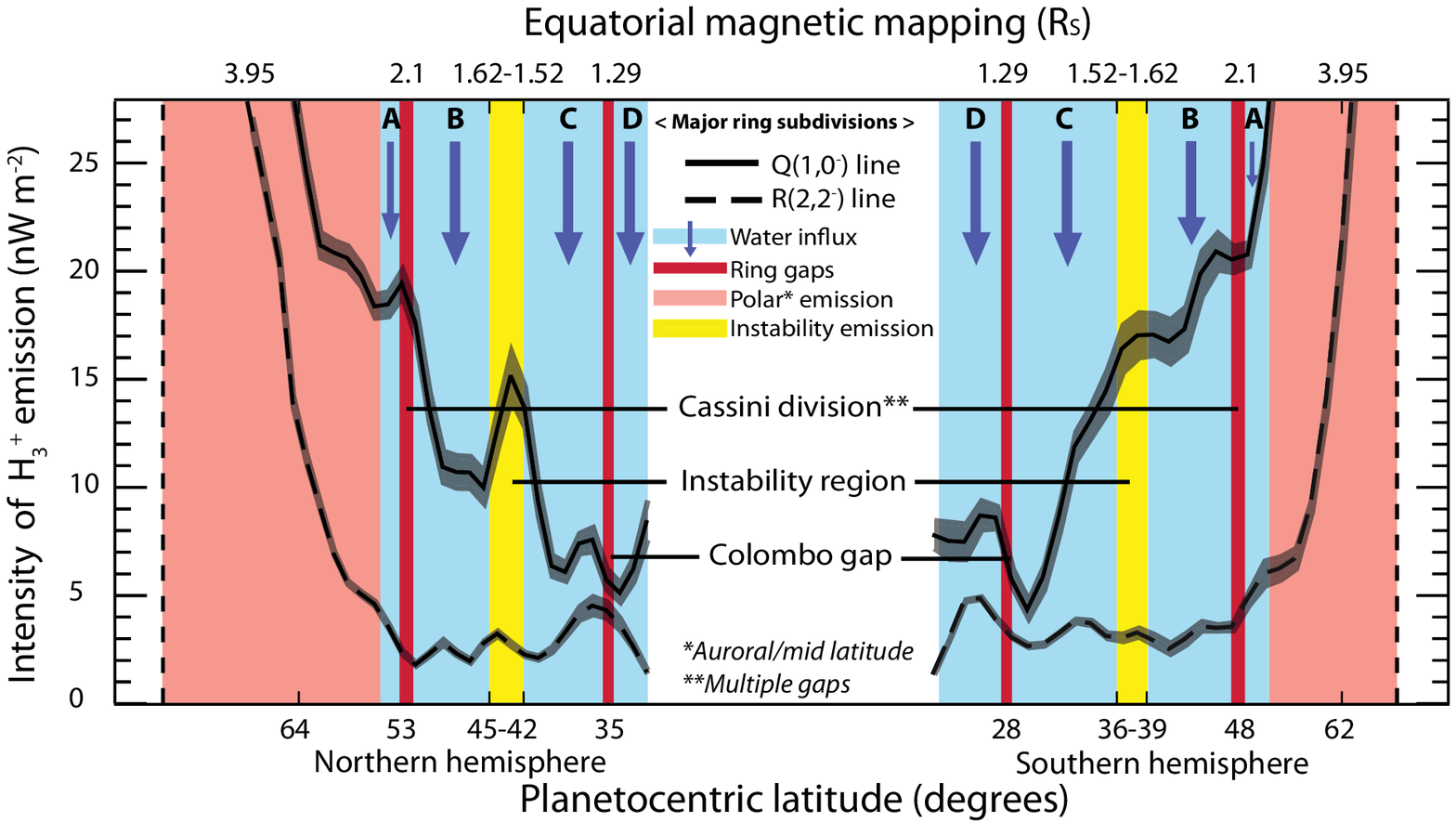}
\caption{{\small \textbf{Intensity of H$_3^+$ IR emission as a function of position along Saturn's noon meridian.} The horizontal axes show a scale of planetocentric latitude at the bottom and the planetocentric equatorial distances which those latitudes magnetically map to at the top, where R$_\mathrm{S}$ is the 1 bar Saturn equatorial radius of 60,268 km. The y-axis shows the intensity of H$_3^+$ emission of the two spectral lines that are shown, Q(1, 0$^-$) at 3.953 $\hbox{\textmu}$m (black line) and R(2, 2$^-$) at 3.622 $\hbox{\textmu}$m (dashed black line) with a central gap where the observed emission is swamped by solar photon reflection from the planetary rings. Latitude bands mapping along planetary magnetic field lines to the main ring subdivisions in the equatorial plane are coloured blue, whilst dark red maps to the ring gaps, as labelled. The yellow shading is the instability region between the stability limits as discussed in the main text. High- to mid-latitude emission is shaded pink out to the limb of the planet (dashed black line), peaking in intensity at $\sim$2 and $\sim$1 $\hbox{\textmu}$W m$^{-2}$ for the Q(1, 0$^-$) and R(2, 2$^-$) lines, respectively. The 1-sigma error in intensity measurements are denoted by the grey shading. The errors in latitude are on average 3$^{\circ}$, mainly caused by the Earth's atmospheric attenuation, i.e. seeing, of 0.4 seconds of arc. To remove additional errors, only the best seeing conditions were selected, such that the Q(1, 0$^-$) line derives from the co-addition of $\sim$40\% of the dataset. However, due to the weaker signal in the R(2, 2$^-$) line, $\sim$90\% of the dataset was co-added - leading to a greater error in latitude but a reduction in intensity errors (see also supplementary information).}}
\end{figure}

\begin{figure}
\includegraphics[width=35pc]{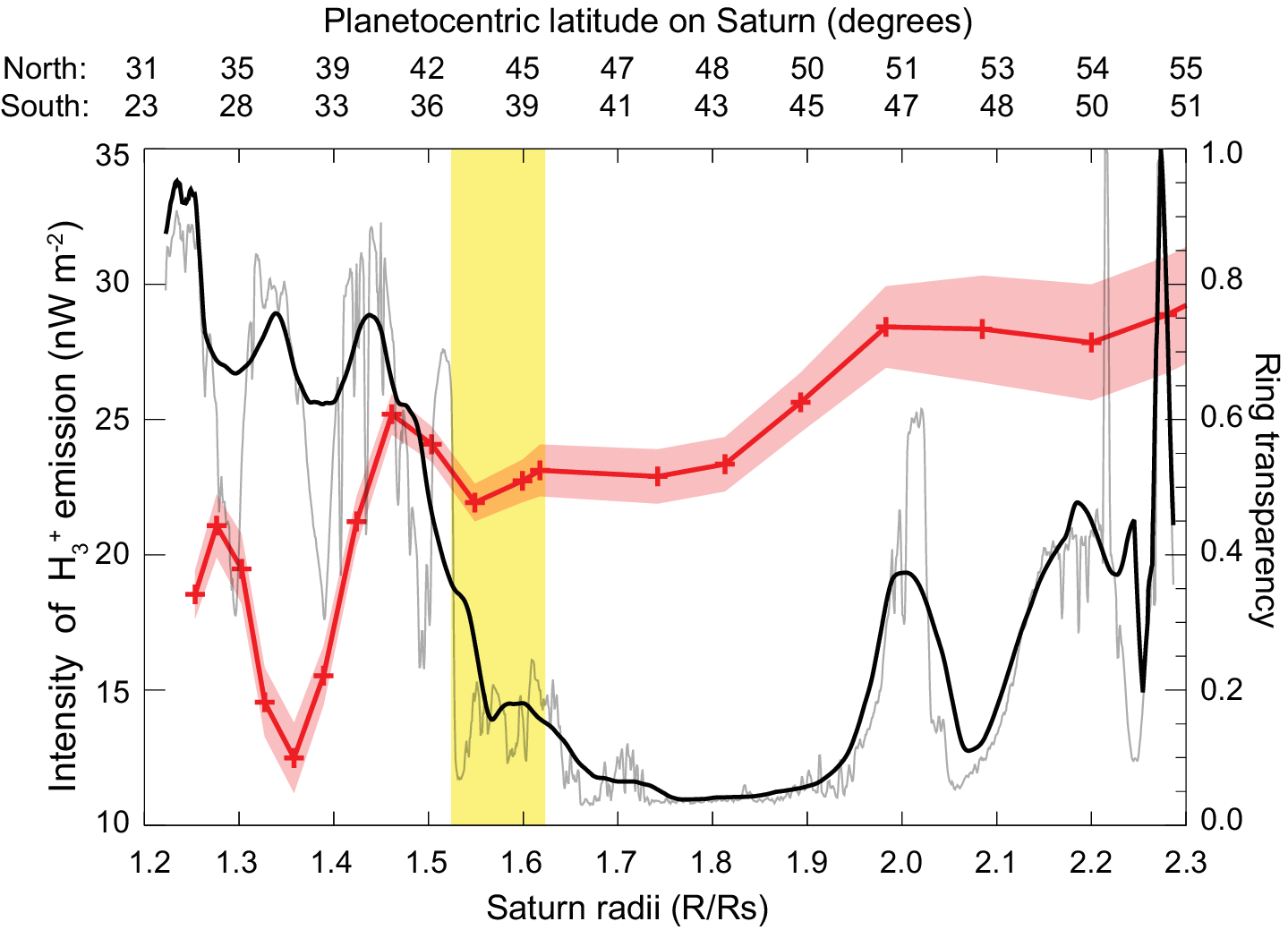}
\caption{\textbf{Comparison of H$_3^+$ intensity and the transparency of Saturn's rings.} The horizontal axes show latitude in each hemisphere at the top in degrees and the corresponding equatorial distance at the bottom, mapped along magnetic field lines. The H$_3^+$ intensity from the Q(1, 0$^-$) and R(2, 2$^-$) lines in both hemispheres are co-added to obtain a high signal-to-noise ratio, and shown by the red line and a line-of-sight correction has been applied. The black and grey lines show a smoothed and raw normalised photon transmission count (transparency), respectively, of the rings taken from Voyager 2 UVS stellar occultation archives\cite{ppsring}. The yellow shading indicates the instability region between the instability radii at 1.52 and 1.62  R$_\mathrm{S}$. The 1-sigma error in the intensity measurements is denoted by the pale red shading. The H$_3^+$ intensity shows peaks in regions of high ring transparency and troughs where transparency is low, indicating that H$_3^+$ emission is quenched when it is mapped to high density regions of the rings. At higher latitudes, one pixel represents more degrees of latitude than at low-latitudes, as depicted by the red crosses which indicate the pixel spacing on the spectrograph CCD.}
\end{figure}

Cassini spacecraft observations over Saturn's rings during its orbit insertion manoeuvre in 2004, unique in the mission to date, indicate the presence of a water-product atmosphere surrounding the rings deriving from the icy grains, that is partly ionised by solar UV analogous to the planetary ionosphere\cite{ringion,2005coates}. Given the correlation between Saturn's ionosphere and magnetically mapped locations in the rings, we propose that it is this charged material that causes the pattern of features seen on the planet. Water-related ions and/or electrons must be driven to stream along the field lines into the planetary ionosphere, and we will now discuss how and in what ways this could produce the modulation seen in the ionosphere. \\

A magnetic link between Saturn's rings and atmosphere has previously been invoke to explain lower than expected electron densities and their latitudinal variations in the planetary ionosphere, through the influx of water\cite{1984conn,1986conn,wilsonring}. These earlier observations show discrete dark bands in the lower atmosphere (beneath the ionosphere) taken by the Voyager 2 spacecraft\cite{1986conn}, which the authors interpreted to be magnetically mapped sequentially to the inner edge of the B ring, the instability radius and the orbital path of Enceladus. Comparing our low-latitude profiles with this past result, we see that water deposited in the lower atmosphere appears to be co-located with the edges of an ionospheric peak extending from 1.52 to 1.62 R$_\mathrm{S}$ in H$_3^+$ emission, within degrees of planetocentric latitude, seen in Fig. 2. \\

Water ions flowing from the rings along magnetic field lines into the ionosphere cause the electron density to be reduced through rapid chemical recombination (quenching)\cite{1986conn}. Water products also deplete H$_3^+$ because it protonates (charge-exchanges) quickly with molecules heavier than H and He\cite{millsept06}, hence a drop in H$_3^+$ density and thus intensity at latitudes where the most water is delivered to the planet should be visible. When mapping along field lines from the ionosphere to the equatorial plane, it is found that each large trough corresponds to a major subdivision of the rings. In the same manner, the peaks in emission are found to map to prominent gaps in the rings. The reason for these peaks in relatively high intensity may be that these are regions in which the influx of water is severely reduced. Thus, these peaks are not really peaks at all, but regions in which the ionosphere is quenched less than latitudes either side of it. 

In the Cassini division, for example, which maps on average to $\sim$2.1 R$_\mathrm{S}$ in Fig. 2, occurs at latitudes where an increase in the emission of H$_3^+$ is clearly visible. Water influx from the A and B ring quenches the ionosphere at locations on either side of these latitudes, leading to the prominent peak seen in-between. This occurs in both hemispheres, symmetrical about the magnetic equator of Saturn. One exception to this apparent correlation between water influx and H$_3^+$ emission is in the ionospheric region mapping between the instability radii. Modelling suggests water influx should peak at these regions\cite{1983hill}, but our measurements clearly show these regions to lie within a peak in emission, rather than at the locations of quenching. A possible explanation for this anomaly is that while significantly enhanced water influx occurs on the edges of this region, there could be low densities of water ion influx, leading to little or no reduction of H$_3^+$ density at latitudes mapping to it. The reduced water source here may be the result of the instability radii consuming the local supply at either side of the instability region, effectively cutting off the supply. However, as stated, the uncertainty in mapping here can affect the interpretation of this peak in emission, so we now explore other mechanisms that could create the observed features. \\

An alternative interpretation for the observations is that the peaks in intensity correspond to temperature increases in H$_3^+$, whilst the troughs correspond to the natural background levels of H$_3^+$ emission produced by solar EUV ionisation. These temperature increases would be the result of Joule heating via the flow of charged particles, which requires that the resultant rise in H$_3^+$ intensity is large enough to overcome any quenching of H$_3^+$ density that may take place. Detailed modelling of the effects of ring rain are required to establish what the background H$_3^+$ emission intensity in these latitudes should be (based on solar EUV ionisation), and whether or not the peaks in emission found here are equal to or higher than this level. The shadow cast by the rings is known to create variations in the ion density with latitude\cite{ringshadow}, but this is unable to explain the features seen here as the shadow falls behind the rings in ground-based geometry; as seen in Fig. 1, the reflection of sunlight by the rings obscures this region entirely.  \\

Atmospheric models still continue to struggle to explain the electron density distribution\cite{1984conn,1986conn,wilsonring} and underestimate the observed temperature of the neutral gases in the upper atmosphere at low-latitudes by many hundreds of Kelvin, in what is known colloquially as the `energy crisis'\cite{YM,2007crisis}. These shortcomings highlight the acute need for a greater understanding of the systems acting upon the low-latitude atmosphere. The observations herein are the first direct measure of the ionosphere's reaction to ring-water input, rather than through measurements of the deeper atmosphere below it. In the past only discrete features linked to water influx have been found, whilst here we see that the scale of this interaction is global, and that water from the A- B- and C- rings can quench the ionosphere, mapping to over 30\% of the planet's surface area. The D-ring mapped latitudes are largely obscured, but in including them, the implication is that up to 43\% of the planet may be subjected to water influx. \\

\textbf{Acknowledgements} \\
The data presented herein were obtained at the W.M. Keck Observatory, which is operated as a scientific partnership among the California Institute of Technology, the University of California and NASA. The observations were made to support the Cassini auroral campaign. Ring profile data were provided by the Planetary Rings Node website\cite{ppsring}. Discussions within the international team led by T.S.S. on `Comparative Jovian Aeronomy' have greatly benefited this work, this was hosted by the International Space Science Institute (ISSI). The UK Science and Technology Facilities Council (STFC) supported this work through the PhD Studentship of J.O'D, and grant support for G.H.J.

\textbf{Author contributions} \\
J.O'D. analysed and interpreted the data and wrote the paper. T.S.S., S.M. and K.B. proposed and designed the study and collected the data. H.M. greatly aided the analysis and interpretation of data. S.W.H.C. provided the magnetic-mapping model and magnetospheric information. G.H.J. provided ring-plane information. J.S.D.B. provided context from Cassini VIMS observations. All authors assisted in the interpretation of data and commented on the manuscript.


\begin{thebibliography}{10}
\expandafter\ifx\csname url\endcsname\relax
  \def\url#1{\texttt{#1}}\fi
\expandafter\ifx\csname urlprefix\endcsname\relax\def\urlprefix{URL }\fi
\providecommand{\bibinfo}[2]{#2}
\providecommand{\eprint}[2][]{\url{#2}}

\bibitem{tssras}
\bibinfo{author}{{Stallard}, T.~S.} \emph{et~al.}
\newblock \bibinfo{title}{{Temperature changes and energy inputs in giant
  planet atmospheres: what we are learning from H$_3^+$}}.
\newblock \emph{\bibinfo{journal}{Phil. Trans. Roy. Soc.}}
  \textbf{\bibinfo{volume}{370}}, \bibinfo{pages}{5213--5224}
  (\bibinfo{year}{2012}).

\bibitem{2010mill}
\bibinfo{author}{{Miller}, S.}, \bibinfo{author}{{Stallard}, T.},
  \bibinfo{author}{{Melin}, H.} \& \bibinfo{author}{{Tennyson}, J.}
\newblock \bibinfo{title}{{H$_3^+$ cooling in planetary atmospheres}}.
\newblock \emph{\bibinfo{journal}{Faraday Discussions}}
  \textbf{\bibinfo{volume}{147}}, \bibinfo{pages}{283--291}
  (\bibinfo{year}{2010}).

\bibitem{1984book}
\bibinfo{author}{{Atreya}, S.~K.}, \bibinfo{author}{{Donahue}, T.~M.},
  \bibinfo{author}{{Nagy}, A.~F.}, \bibinfo{author}{{Waite}, J.~H., Jr.} \&
  \bibinfo{author}{{McConnell}, J.~C.}
\newblock \emph{\bibinfo{title}{{Theory, measurements, and models of the upper
  atmosphere and ionosphere of Saturn}}}, \bibinfo{pages}{239--277}
  (\bibinfo{year}{1984}).

\bibitem{2007crisis}
\bibinfo{author}{{Smith}, C.~G.~A.}, \bibinfo{author}{{Aylward}, A.~D.},
  \bibinfo{author}{{Millward}, G.~H.}, \bibinfo{author}{{Miller}, S.} \&
  \bibinfo{author}{{Moore}, L.~E.}
\newblock \bibinfo{title}{{An unexpected cooling effect in Saturn's upper
  atmosphere}}.
\newblock \emph{\bibinfo{journal}{Nature}} \textbf{\bibinfo{volume}{445}},
  \bibinfo{pages}{399--401} (\bibinfo{year}{2007}).

\bibitem{2010moore}
\bibinfo{author}{{Moore}, L.}, \bibinfo{author}{{Mueller-Wodarg}, I.},
  \bibinfo{author}{{Galand}, M.}, \bibinfo{author}{{Kliore}, A.} \&
  \bibinfo{author}{{Mendillo}, M.}
\newblock \bibinfo{title}{{Latitudinal variations in Saturn's ionosphere:
  Cassini measurements and model comparisons}}.
\newblock \emph{\bibinfo{journal}{J. Geophys. Res.}}
  \textbf{\bibinfo{volume}{115}}, \bibinfo{pages}{11317}
  (\bibinfo{year}{2010}).

\bibitem{1984conn}
\bibinfo{author}{{Connerney}, J.~E.~P.} \& \bibinfo{author}{{Waite}, J.~H.}
\newblock \bibinfo{title}{{New model of Saturn's ionosphere with an influx of
  water from the rings}}.
\newblock \emph{\bibinfo{journal}{Nature}} \textbf{\bibinfo{volume}{312}},
  \bibinfo{pages}{136--138} (\bibinfo{year}{1984}).

\bibitem{1986conn}
\bibinfo{author}{{Connerney}, J.~E.~P.}
\newblock \bibinfo{title}{{Magnetic connection for Saturn's rings and
  atmosphere}}.
\newblock \emph{\bibinfo{journal}{Geophys. Res. Lett.}}
  \textbf{\bibinfo{volume}{13}}, \bibinfo{pages}{773--776}
  (\bibinfo{year}{1986}).

\bibitem{wilsonring}
\bibinfo{author}{{Wilson}, G.~R.} \& \bibinfo{author}{{Waite}, J.~H., Jr.}
\newblock \bibinfo{title}{{Kinetic modeling of the Saturn ring-ionosphere
  plasma environment}}.
\newblock \emph{\bibinfo{journal}{J Geophys. Res.}}
  \textbf{\bibinfo{volume}{94}}, \bibinfo{pages}{17287--17298}
  (\bibinfo{year}{1989}).

\bibitem{nirs}
\bibinfo{author}{{McLean}, I.~S.} \emph{et~al.}
\newblock \bibinfo{title}{{Design and development of NIRSPEC: a near-infrared
  echelle spectrograph for the Keck II telescope}}.
\newblock In \bibinfo{editor}{{Fowler}, A.~M.} (ed.)
  \emph{\bibinfo{booktitle}{Society of Photo-Optical Instrumentation Engineers
  (SPIE) Conference Series}}, vol. \bibinfo{volume}{3354} of
  \emph{\bibinfo{series}{Society of Photo-Optical Instrumentation Engineers
  (SPIE) Conference Series}}, \bibinfo{pages}{566--578} (\bibinfo{year}{1998}).

\bibitem{1982hill}
\bibinfo{author}{{Northrop}, T.~G.} \& \bibinfo{author}{{Hill}, J.~R.}
\newblock \bibinfo{title}{{Stability of negatively charged dust grains in
  Saturn's ring plane}}.
\newblock \emph{\bibinfo{journal}{J. Geophys. Res.}}
  \textbf{\bibinfo{volume}{87}}, \bibinfo{pages}{6045--6051}
  (\bibinfo{year}{1982}).

\bibitem{1983hill}
\bibinfo{author}{{Northrop}, T.~G.} \& \bibinfo{author}{{Hill}, J.~R.}
\newblock \bibinfo{title}{{The inner edge of Saturn's B ring}}.
\newblock \emph{\bibinfo{journal}{J. Geophys. Res.}}
  \textbf{\bibinfo{volume}{88}}, \bibinfo{pages}{6102--6108}
  (\bibinfo{year}{1983}).

\bibitem{burt2010}
\bibinfo{author}{{Burton}, M.~E.}, \bibinfo{author}{{Dougherty}, M.~K.} \&
  \bibinfo{author}{{Russell}, C.~T.}
\newblock \bibinfo{title}{{Saturn's internal planetary magnetic field}}.
\newblock \emph{\bibinfo{journal}{J. Geophys. Res.}}
  \textbf{\bibinfo{volume}{37}}, \bibinfo{pages}{24105} (\bibinfo{year}{2010}).

\bibitem{ringion}
\bibinfo{author}{{Luhmann}, J.~G.}, \bibinfo{author}{{Johnson}, R.~E.},
  \bibinfo{author}{{Tokar}, R.~L.}, \bibinfo{author}{{Ledvina}, S.~A.} \&
  \bibinfo{author}{{Cravens}, T.~E.}
\newblock \bibinfo{title}{{A model of the ionosphere of Saturn's rings and its
  implications}}.
\newblock \emph{\bibinfo{journal}{Icarus}} \textbf{\bibinfo{volume}{181}},
  \bibinfo{pages}{465--474} (\bibinfo{year}{2006}).

\bibitem{2005coates}
\bibinfo{author}{{Coates}, A.~J.} \emph{et~al.}
\newblock \bibinfo{title}{{Plasma electrons above Saturn's main rings: CAPS
  observations}}.
\newblock \emph{\bibinfo{journal}{Geophys. Res. Lett.}}
  \textbf{\bibinfo{volume}{32}}, \bibinfo{pages}{14} (\bibinfo{year}{2005}).

\bibitem{millsept06}
\bibinfo{author}{{Miller}, S.}, \bibinfo{author}{{Stallard}, T.},
  \bibinfo{author}{{Smith}, C.} \& \bibinfo{author}{{et al.}}
\newblock \bibinfo{title}{{H$_3^+$: the driver of giant planet atmospheres}}.
\newblock \emph{\bibinfo{journal}{Phil. Trans. Roy. Soc. London.}}
  \textbf{\bibinfo{volume}{364}}, \bibinfo{pages}{3121--3137}
  (\bibinfo{year}{2006}).

\bibitem{ringshadow}
\bibinfo{author}{{Mendillo}, M.} \emph{et~al.}
\newblock \bibinfo{title}{{Effects of ring shadowing on the detection of
  electrostatic discharges at Saturn}}.
\newblock \emph{\bibinfo{journal}{Geophys. Res. Lett.}}
  \textbf{\bibinfo{volume}{32}}, \bibinfo{pages}{5107} (\bibinfo{year}{2005}).

\bibitem{YM}
\bibinfo{author}{Yelle, R.~V.} \& \bibinfo{author}{Miller, S.}
\newblock \emph{\bibinfo{title}{Jupiter’s upper atmosphere. In Jupiter: the
  planet, satellites and magnetosphere}}, \bibinfo{pages}{185--218}
  (\bibinfo{publisher}{Cambridge University Press.},
  \bibinfo{address}{Cambridge, UK}, \bibinfo{year}{2004}).

\bibitem{ppsring}
\bibinfo{author}{{Lillie}, C.~F.}, \bibinfo{author}{{Hord}, C.~W.},
  \bibinfo{author}{{Pang}, K.}, \bibinfo{author}{{Coffeen}, D.~L.} \&
  \bibinfo{author}{{Hansen}, J.~E.}
\newblock \bibinfo{title}{{The Voyager mission photopolarimeter experiment}}.
\newblock \emph{\bibinfo{journal}{Space Sci. Rev.}}
  \textbf{\bibinfo{volume}{21}}, \bibinfo{pages}{159--181}
  (\bibinfo{year}{1977}).

\end{thebibliography}
\end{document}